\documentclass[11pt,twoside]{article}
\usepackage{asp2006}
\usepackage{graphicx}
\begin{document}
\title{Understanding the Spins of Young Stars}
\author{Sean Matt}
\affil{Department of Astronomy, University of Virginia, Charlottesville VA 22904, USA}
\author{Ralph E. Pudritz}
\affil{Department of Physics \& Astronomy, McMaster University, Hamilton ON L8S 4M1, Canada}

\begin{abstract}
We review the theoretical efforts to understand why pre-main-sequence
stars spin much more slowly than expected.  The first idea put forward
was that massive stellar winds may remove substantial angular
momentum.  Since then, it has become clear that the magnetic
interaction between the stars and their accretion disks explains many
of the observed emission properties.  The disk locking scenario, which
assumes the magnetic star-disk interaction also solves the stellar
spin problem, has received the most attention in the literature to
date.  However, recent considerations suggest that the torques in the
star-disk interaction are insufficient for disk locking to explain the
slow rotators.  This prompts us to revisit stellar winds, and we
conclude that stellar winds, working in conjunction with
magnetospheric accretion, are a promising candidate for solving the
angular momentum problem.  We suggest future directions for both
observations and theory, to help shed light on this issue.
\end{abstract}
\section{Introduction: The Angular Momentum Problem for Young Stars}

Before the rotation rates of optically visible pre-main-sequence stars
(T Tauri stars) were known, it was expected that these stars would
spin at or very near breakup speed \citep{vogelkuhi81}.  This
expectation was well-founded, since molecular cloud cores, which
collapse to form stars, are observed to have $\sim 10^4$ times more
angular momentum than a star rotating at breakup speed.  Thus,
collapse proceeds to first form a Keplerian disk, and stars are built
up by accreting material from this disk.  Even if a protostar was born
with slow spin, the accretion of this disk material with high specific
angular momentum would spin up the star.  The timescale for an
accreting star to spin up by some amount, $\Delta f$, expressed as a
fraction of breakup speed, is given by
\begin{eqnarray}
\label{eqn_tsu}
\tau_{\rm su} \approx 10^5 {\rm yrs} 
      \left({\Delta f \over 0.1}\right)
      \left({k^2 \over 0.2}\right)
      \left({{{R_{\rm t} / R_*} \over {3}}}\right)^{-1/2}
      \left({M_* \over 0.5 M_\odot}\right)
      \left({\dot M_{\rm a} \over 
	5 \times 10^{-8} M_\odot / {\rm yr}}\right)^{-1},
\end{eqnarray}
where $k^2$ is the square of the mean radius of gyration (defined by
the moment of inertia $I_* = k^2 M_* R_*^2$); $R_*$ and $M_*$ are the
stellar radius and mass; $R_{\rm t}$ is the radius of the disk inner
edge; $\dot M_{\rm a}$ is the mass accretion rate; and $f = \Omega_*
R_*^{3/2} (G M_*)^{-1/2}$, where $\Omega_*$ is the angular rotation
rate of the star.  Thus, a typical T Tauri star should spin up by
$\approx 10$\% of breakup speed every hundred thousand years.  Since
protostars are not likely born with slow spin, and since T Tauri ages
are typically several times $10^{5-6}$ yrs, one expects them to rotate
rapidly.

Adding to the expectation of rapid rotation, pre-main-sequence stars
contract by a factor of several times before they reach the main
sequence.  Furthermore, it is well known that main sequence stars with
spectra type later than $\sim$F2 lose angular momentum via magnetized,
coronal stellar winds.  But the timescale for this spin down is of the
order of billions of years.  Since the pre-main-sequence phase lasts
only a few tens of millions of years, an ordinary (main-sequence-like)
stellar wind cannot be important for angular momentum evolution of a T
Tauri star.

However, over the last few decades, measurements of $v \sin i$ and
rotation periods for large samples of T Tauri stars \citep[][see
\citealp{rebull3ea04} for a compilation]{vogelkuhi81, herbstea07} have
revealed that approximately half of the stars are rotating at only
$\sim 10$\% of breakup speed.  How do these stars rid themselves of
angular momentum?

\section{Stellar Winds, Take One} \label{sec_sw1}

The first proposed solution to this problem, by
\citet{hartmannstauffer89}, was that substantial angular momentum
could be removed by a magnetized stellar wind \citep[see
also][]{mestel84, toutpringle92}.  These winds would have to be
massive, with mass outflow rates comparable to the accretion rates,
and the stars should be more highly magnetized than their main
sequence counterparts.  \citet{hartmannea90} computed optical emission
line profiles (of hydrogen, Mg {\small II}, Ca {\small II}, and Na
{\small I}) expected to arise in such a wind, with a temperature $\sim
10^4$ K.  However, these computed lines, which typically exhibit broad
P-Cygni profiles, did not match very well with those observed.

As discussed in the next section, it is now clear that those lines
arise primarily from material accreting onto the star
\citep{muzerolle3ea01}, though a small contribution to the flux from a
disk wind or stellar wind might be present \citep{kurosawa3ea06}.
Furthermore, as discussed in section 5, the stellar winds are likely
to be much hotter than $10^4$ K, so the emission from the wind may be
quite different than that considered by \citet{hartmannea90}.

\section{The Magnetic Star-Disk Interaction}

Around the same time as the work of \citet{hartmannea90},
\citet{camenzind90} and \citet{konigl91} applied the magnetic
accretion model of \citet{ghoshlamb78} to T Tauri stars.  This model
was originally developed for accreting neutron stars, and it neglects
any effects of a stellar wind.  A common feature of all magnetic
star-disk interaction models is that the stellar magnetosphere is
strong enough to disrupt the accretion disk at some radius ($R_{\rm
t}$) above the surface of the star.  From there, accreting material is
channeled by the magnetic field lines onto the stellar surface at near
free-fall velocities.

The truncation of the disk occurs where the stellar magnetosphere is
able to spin down the disk material so that it is no longer
centrifugally supported against gravity.  Thus, the star feels a spin
up torque due to the disk truncation and subsequent accretion of
material equal to the accretion rate times the specific angular
momentum of material rotating at Keplerian speed at $R_{\rm t}$.  This
accretion torque is
\begin{eqnarray}
\label{eqn_tacc}
\dot J_{\rm a} = \dot M_{\rm a} \sqrt{G M_* R_{\rm t}}.
\end{eqnarray}
We used this torque to calculate the spin up time in equation
\ref{eqn_tsu}.

This basic concept of disk truncation and magnetospheric accretion has
been very successful at explaining the observed line profiles and
fluxes (e.g., of hydrogen, Ca {\small II}, Na D) as arising from
material that is flowing along the magnetic field line onto the
surface of the star \citep[e.g.][]{muzerolle3ea01}.  Also, shocks
formed by accreting material striking the stellar surface provides a
natural explanation for observed uv excesses and hot spots on the
surface of the T Tauri stars \citep{konigl91}.  It is now clear that
magnetospheric accretion is indeed an important process occuring in
accreting T Tauri stars.  However, it is important to note that the
observations supporting magnetospheric accretion do not address the
angular momentum flow in the star-disk interaction, which is the topic
of the remainder of this section.

     \subsection{Disk Locking}

It is possible for some stellar magnetic flux to connect with the disk
beyond the truncation radius, $R_{\rm t}$.  Due the the differential
rotation between the star and disk, the magnetic field is twisted
azimuthally and imparts a torque on the star.  At the corotation
radius, $R_{\rm co} = f^{-2/3} R_*$, the star and disk rotate at the
same angular rate.  Assuming Keplerian rotation in the disk, the
magnetic flux that connects the star to the region in the disk between
$R_{\rm t}$ and $R_{\rm co}$ will act to spin up the star, while the
flux connecting to the disk outside $R_{\rm co}$ will remove angular
momentum from the star.

In order to address the angular momentum problem, the
\citet{ghoshlamb78} model assumes that the stellar dipole magnetic
field connects to the surface of the disk over a large range of radii.
This provides a possible explanation for angular momentum loss, as
long as there is a substantial amount of stellar magnetic flux that
connects to the disk outside $R_{\rm co}$, and $R_{\rm t}$ needs to be
very close to $R_{\rm co}$.  This way, the torque on the star
associated with the magnetic connection to the disk can be negative
and possibly strong enough to counteract the spin up torque from
accretion (eq.\ \ref{eqn_tacc}).

Adopting the Ghosh \& Lamb framework, a few authors
\citep{cameroncampbell93, armitageclarke96, pinzon06} have followed
the spin evolution of pre-main-sequence stars according to the angular
momentum equation
\begin{eqnarray}
\label{eqn_angmom}
I_* {\partial \Omega_* \over \partial t} = 
     - \Omega_* {\partial I_* \over \partial t} +
     \dot J_{\rm a} + \dot J_{\rm m},
\end{eqnarray}
where $\dot J_{\rm m}$ is the total torque due to the twisting of the
connected stellar magnetic flux.  Stellar models provide information
about the rate of change of the stellar moment of inertia, $\partial
I_* / \partial t$.  Regardless of the choice of initial stellar spin
rate, these authors find that, under most circumstances, the system
evolves to a state in which the net torque on the star is very nearly
zero within a timeframe of a few hundred thousand years (as expected
from eq.\ \ref{eqn_tsu}).

The spin rate of the star in this torque equilibrium state can be
calculated directly from the Ghosh \& Lamb model, which gives
\begin{eqnarray}
\label{eqn_dlspin}
\Omega_*^{\rm eq} = C \dot M_{\rm a}^{3/7} (G M_*)^{5/7} \mu^{-6/7},
\end{eqnarray}
where $C$ is a constant (``fudge factor'') of order unity, $\mu \equiv
B_* R_*^3$ is the dipole moment, and $B_*$ is the magnetic field
strength at the surface of the star.  In this state, the truncation
radius must be very close to the corotation radius.  Thus, the stellar
rotation rate is very nearly the same as that at the disk inner edge,
and this is referred to as the ``disk locked'' state.

There are several models in the literature that follow similar
assumptions as the Ghosh \& Lamb model.  Each of these differ in the
details of how they treat the magnetic coupling between the stellar
field and the disk \citep[see][]{mattpudritz05}.  In the end, they all
derive the same formula (\ref{eqn_dlspin}), but with slightly
different values of the factor $C$ of order unity.  For example,
\citet{konigl91} used $C \approx 1.15$.

The X-wind \citep[][and subsequent works]{shuea94} is a disk locking
model that removes the excess angular momentum by means of a wind from
the inner edge of the disk.  To do this, the X-wind assumes the
stellar dipole magnetic field lines are pinched and connect to a very
small region (the ``X-point'') around $R_{\rm co}$, rather than
connecting to a large portion of the disk.  Also, the X-wind assumes a
torque equilibrium state, rather than producing a general formulation
for a spin up or spin down state.  Because it is a disk locking model,
the X-wind predicts the same equilibrium spin as in equation
(\ref{eqn_dlspin}), with its own factor $C$ of order unity.
\citet{ostrikershu95} found $C \approx 1.13$.

We can use equation (\ref{eqn_dlspin}) to predict the stellar dipole
field strength necessary to explain the existence of slow rotators,
\begin{eqnarray}
\label{eqn_bfield}
B_*^{\rm eq}  & \approx & 900 ~{\rm Gauss} 
      \left({C \over 1.0}\right)^{7/6} 
      \left({f \over 0.1}\right)^{-7/6} \times  \nonumber \\ 
      & &\left({{R_*} \over {2 R_\odot}}\right)^{-5/4}
      \left({M_* \over 0.5 M_\odot}\right)^{1/4}
      \left({\dot M_{\rm a} \over 
	5 \times 10^{-8} M_\odot / {\rm yr}}\right)^{1/2},
\end{eqnarray}
It is evident that the Ghosh \& Lamb-type and the X-wind disk locking
models require stellar field strengths of the order of a kilo-Gauss
\citep[eq.\ \ref{eqn_bfield}; see][]{johnskrull3ea99}.  Furthermore,
disk locking always requires a large amount of stellar flux to connect
to the disk.

     \subsection{Problems With Disk Locking}

\begin{figure}[!ht]
\begin{center}
\includegraphics{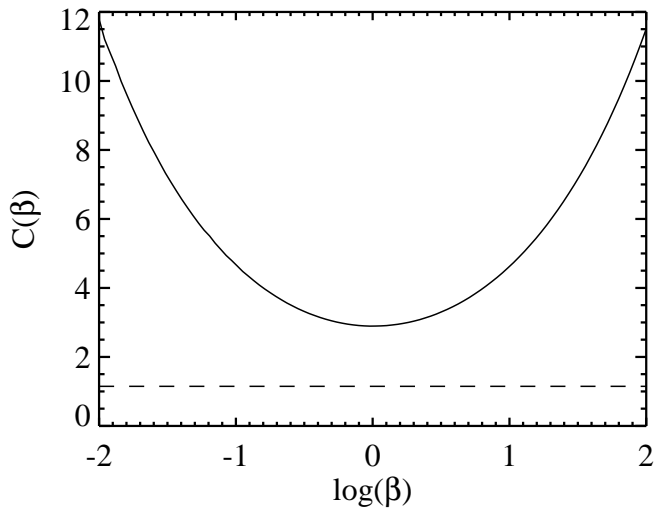}
\end{center}
\caption{The factor $C$ in equation (\ref{eqn_dlspin}) as a function
of the magnetic coupling parameter $\beta$ quantifies the effect of
field line opening.  The dashed line corresponds to the value of $C =
1.15$ used by \citet{konigl91}.  The figure is adapted from
\citet{mattpudritz05}.}\label{fig_cbg}
\end{figure}

Kilo-Gauss fields were in fact a prediction of disk locking
\citep{konigl91}. So when kilo-Gauss strength mean fields were first
observed in T Tauri stars \citep[e.g.,][]{basri3ea92, guenther97,
johnskrull3ea99}, this seemed to be evidence for disk locking.
However, these magnetic field measurements do not probe the global,
dipole field strength \citep[e.g.,][]{safier98, johnskrullea99},
required for efficient angular momentum loss.  Rather,
spectropolarimetric observations of a handful of accreting T Tauri
stars to date (see contribution by Johns-Krull in these proceedings)
reveal upper limits or marginal detections of the global field of
$\sim 100$ Gauss.  This is a serious problem for understanding the
existence of the slow rotators, since for these field strengths,
equation (\ref{eqn_dlspin}) predicts spin rates in excess of 50\% of
breakup.

A second and independent problem with the disk locking scenario
regards the need for lots of stellar flux to connect to the disk.
\citet{safier98} pointed out that stellar winds should open
(disconnect from the disk) all of the stellar flux reaching outside
$\sim 3 R_*$, affecting all disk locking models.  Alternatively,
several authors \citep{lyndenbellboily94, lovelace3ea95,
agapitoupapaloizou00, uzdensky3ea02} showed that differential rotation
between the star and disk will lead to an opening of much of the field
that is assumed closed in the Ghosh \& Lamb model.  In particular,
\citet{uzdensky3ea02} showed that the magnetic connection between the
star and disk becomes severed when the magnetic field is twisted to a
point where $B_\phi / B_z$ is greater than one, where $B_z$ is the
vertical vector component of the field, and $B_\phi$ is the azimuthal
component generated by the twisting.

This amount of twisting occurs in approximately one half orbit of the
disk, indicating that a large scale magnetic connection between the
star and disk cannot persist for long.  The only way for the magnetic
field to remain connected to the disk is for the magnetic field to
``slip'' through the disk (e.g., via turbulent diffusion or magnetic
reconnection) at a rate equal to the differential rotation rate, so
that $B_\phi / B_z$ remains small.  The slip rate of the magnetic
field relative to the disk depends on the physics of the magnetic
coupling to the disk.  \citet{mattpudritz05} characterized this
coupling with a parameter $\beta$, such that the slip rate of the
magnetic field through the disk equals $\beta (B_\phi / B_z) v_{\rm
K}$, where $v_{\rm K}$ is the Keplerian speed.  Thus, large $\beta$
corresponds to weak magnetic coupling (fast slipping), and small
$\beta$ means strong coupling.

\citet{mattpudritz05} quantified the effect that the field opening has
on the torque felt by the star.  To do this, they followed the
assumptions of \citet{ghoshlamb78} except that the torque was set to
zero where $B_\phi / B_z$ exceeds unity, corresponding to where the
magnetic field is expected to open.  They were able to derive equation
({\ref{eqn_dlspin}) such that the factor $C$ is a function of $\beta$
and contains all the effects of field opening.  The solid line in
figure \ref{fig_cbg} shows this factor $C$ as a function of $\log
\beta$.  For reference, the dashed line indicates the value of $C$
used by \citet{konigl91}.

The behavior of $C(\beta)$ can be understood as follows.  In the
strong coupling limit ($\beta \ll 1$), the field becomes more open as
the coupling gets stronger; so the spin-down torque decreases with
$\beta$, and this means the equilibrium spin rate increases.  In the
weak coupling limit ($\beta \gg 1$), the magnetic field is more
connected, but the slip rate of the field is faster; so the field is
less twisted for increasing $\beta$, and so the spin-down torque
decreases with increasing $\beta$.  Assuming the slipping or diffusion
of the magnetic field is comparable to the effective viscosity in the
disk \citep[e.g., in an $\alpha$-disk;][]{shakurasunyaev73},
\citet{mattpudritz05} estimated that the likely value of $\beta$ in T
Tauri systems is $\sim 10^{-2}$.  From figure \ref{fig_cbg} and
equation (\ref{eqn_dlspin}), this value of $\beta$ predicts an
equilibrium spin rate more than 10 times faster than predicted by the
Ghosh \& Lamb model.  \citet{mattpudritz05} concluded that, when the
effect of the field opening is taken into account, the slowly rotating
T Tauri stars cannot be explained by a disk locking scenario.

\section{Stellar Winds, Reincarnated}

\begin{figure}[!ht]
\begin{center}
\includegraphics{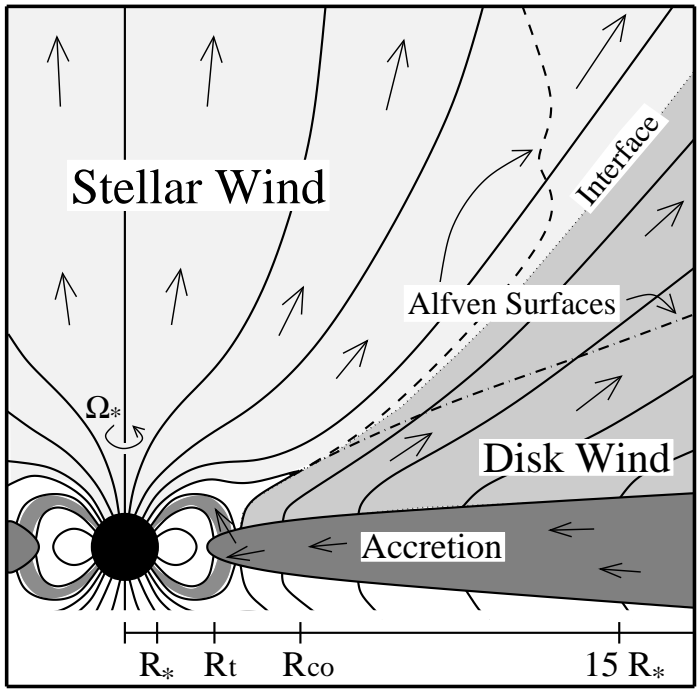}
\end{center}
\caption{Matter and angular momentum flow in the star-disk
interaction.  From \citet{mattpudritz05b}.}\label{fig_cartoon}
\end{figure}

The problems with disk locking, and the expectation that the stellar
magnetic field will be largely open, prompts us to to revisit the
suggestion by \citet{hartmannstauffer89} that stellar winds from T
Tauri stars can remove substantial angular momentum.  Since stellar
winds were first proposed, however, we have gained a more detailed
picture of the star-disk interaction.  In particular, we now know that
much of the emission properties of T Tauri stars arise from
magnetospheric accretion phenomena.  So now a more complete picture of
the interaction between the star and disk, including the transport of
angular momentum by a stellar wind, is presented in figure
\ref{fig_cartoon}.

In the simplest case, the star is magnetically connected only to the
inner edge of the disk and experiences only a spin up torque from this
interaction (namely, the accretion torque, eq.\ \ref{eqn_tacc}).  The
stellar wind extracts angular momentum at a rate
\citep[e.g.][]{weberdavis67, mestel84}
\begin{eqnarray}
\label{eqn_tw}
\dot J_{\rm w} = - \dot M_{\rm w} \Omega_* r_{\rm A}^2,
\end{eqnarray}
where $\dot M_{\rm w}$ is the mass outflow rate in the stellar wind
alone and $r_{\rm A}$ is the radius where the flow speed equals the
magnetic Alfv\'en wave speed.  Note that, strictly speaking, the wind
is 3-dimensional, so the value of $r_{\rm A}^2$ in equation
(\ref{eqn_tw}) actually represents the mass-loss weighted average of
$r_{\rm A}^2$ over a 3-dimensional Alfv\'en surface.

Assuming the system will evolve to a state with a very low net torque,
we can combine equations (\ref{eqn_tacc}) and (\ref{eqn_tw}) to find
the equilibrium spin rate \citep{mattpudritz05b}, expressed as a
fraction of breakup speed,
\begin{eqnarray}
\label{eqn_feq}
f_{\rm eq} \approx 0.1
      \left({{{r_{\rm A} / R_*} \over {12}}}\right)^{-2}
      \left({{{\dot M_{\rm w}/\dot M_{\rm a}} \over {0.1}}}\right)^{-1}
      \left({{{R_{\rm t} / R_*} \over {3}}}\right)^{1/2}.
\end{eqnarray}
So the existence of slow rotators can be explained, if the Alfv\'en
radius is comparable to the solar value \citep[for the sun $r_{\rm A}
\approx 12 R_*$;][]{li99}, and the mass outflow rate in the stellar
wind is a substantial fraction of the accretion rate.  The latter
requires a lot of energy to drive the wind.  \citet{mattpudritz05b}
suggested the wind is driven energetically by accretion power and
showed that of the order of 10\% of the accretion power is needed.

This scenario holds that substantial angular momentum loss occurs only
while the stellar wind is significantly enhanced by accretion power.
Thus there should be a link between a slow equilibrium spin and
accretion \citep[proposed by][]{edwardsea93}.  \citet[][and see
contribution by Rebull in these proceedings]{rebull3ea04} showed that
such a link might account for the evolution of the distribution of T
Tauri star spins, observed to vary from cluster to cluster.

\section{The Nature of Accretion Powered Stellar Winds} \label{sec_nature}

It seems now that, after being largely neglected for $\sim 15$ years,
a stellar wind may still be a promising candidate for solving the
angular momentum problem for young stars.  However, more theoretical
work and observations will be necessary to understand the true nature
and importance of T Tauri star winds.  Here is what we can say thus
far.

Large-scale flows with mass outflow rates comparable to ($\sim 10$\%
of) the accretion rates do emanate from these systems
\citep[e.g.,][]{reipurthbally01}.  It is clear that much of this flow
originates from the disk \citep{blandfordpayne82, shuea95}, but it is
not yet clear what fraction of the total flow can originate from the
star.  The best estimate to date is from \citet{decampli81}, who
showed that the outflow rates of stellar coronal winds with
temperatures of $\approx 10^6$ K cannot exceed $10^{-9} M_\odot$
yr$^{-1}$, or else the X-ray luminosity from the wind would exceed
observed values.  For now, we simply emphasize that a wind from the
disk provides the best explanation for the bulk of the large-scale
outflow, while the primary importance of the stellar wind may be to
remove angular momentum from the star.  Figure \ref{fig_cartoon}
illustrates both wind components.

T Tauri stars are magnetically active and possess hot coronae that are
4--5 orders of magnitude more energetic than the solar corona
\citep{feigelsonmontmerle99}.  Therefore, it seems reasonable to
assume that T Tauri stars possess a thermally driven wind, like a
scaled-up (in $\dot M_{\rm w}$ and $B_*$) solar coronal wind.  The
wind temperature required for thermal driving scales proportional to
the square of the escape speed from the star \citep{parker58}.  This
means that for a typical T Tauri star with $M_* = 0.5 M_\odot$ and
$R_* = 2 R_\odot$, the temperature at the base of the stellar wind
should be $\approx 300,000$ K.  This is much cooler than the $10^6$ K
used by \citet{decampli81}, and so the mass outflow rate can be much
higher than $10^{-9} M_\odot$ yr$^{-1}$ before the X-ray emission
becomes a problem.  We have computed a number of magnetic coronal
winds for T Tauri stars and found viable solutions to equation
(\ref{eqn_feq}) to explain the slow rotators (Matt \& Pudritz 2007, in
preparation).  The way in which accretion power transfers to heat the
stellar corona is yet unknown, though it likely would involve the
dissipation of magneto-acoustic waves and/or mixing of shock-heated
accreting gas into the stellar wind.

\citet{edwardsea03, edwardsea06} observed the He I 10830 line in
several accreting T Tauri stars.  In many cases, they found broad
P-Cygni profiles similar to the hydrogen line profiles predicted by
the stellar wind models of \citet[][see \S 2]{hartmannea90}.  A
stellar wind, as opposed to a disk wind, provides the best explanation
for the large range of blue-shifted velocities over which the line is
absorbed.  \citet{dupreeea05} observed the He I 10830 line, as well as
uv lines of C {\small III} and O {\small VI}, in two systems and
concluded that the stellar wind had a temperature of $\sim 300,000$ K.
Observations such as these are important for our understanding of the
nature of T Tauri stellar winds.

\section{Challenges}

Here we list just a few of the most important outstanding questions
for understanding the angular momentum loss in young stars.

What are the mass outflow rates of the stellar winds (as distinct from
the disk wind) in accreting T Tauri stars?  What is the temperature of
these winds?  For this, radiative transfer modeling including stellar
winds, disk winds, and accretion flow will be useful.  Also, magnetic
field measurements that gauge the strength of the global (dipole)
component of the field are crucial to quantify the strength of
magnetic torques.

Can we understand the observed distributions of T Tauri star spins,
and evolution of the distributions, in the context of an
accretion-powered stellar wind model, in a similar way as has been
done for the disk locking scenario (as discussed by Rebull, these
proceedings)?  How does accretion power transfer to the stellar wind?
Is it thermally driven or otherwise?  What emission properties are
expected from an accretion-powered stellar wind?

Answers to many of these questions are already being pursued by a
number of research groups.  We are optimistic that the combination of
precision spectroscopy and advanced numerical simulations and
theoretical work will bring many new insights to the solution of one
of the most interesting and difficult problems in stellar
astrophysics.

\acknowledgments We would like to thank the organizers for a fun and
productive workshop, and this work benefited from discussions with
numerous attendees (including Gibor Basri, Steve Cranmer, Andrea
Dupree, Lee Hartmann, Chris Johns-Krull, Luisa Rebull, Jeff Valenti,
and Sidney Wolff).  Sean Matt acknowledges support from the University
of Virginia through a Levinson/VITA Fellowship, partially funded by
The Frank Levinson Family Foundation through the Peninsula Community
Foundation.



\end{document}